\documentclass{article}
\usepackage{qcircuit}
\usepackage{amssymb}
\usepackage{physics}
\usepackage[colorlinks]{hyperref} 
\usepackage{graphicx}      
\usepackage{amsmath,xparse}

\def\m{\mu}

\def\be{\begin{equation}}
\def\ee{\end{equation}}
\def\ba{\begin{eqnarray}}
\def\ea{\end{eqnarray}}
\def\la{\langle}
\def\ra{\rangle}

\def\be{\begin{equation}}
\def\ee{\end{equation}}
\def\ba{\begin{eqnarray}}
\def\ea{\end{eqnarray}}
\def\lo{\longrightarrow}
\def\h{\hskip 1cm }

\def\la{\langle}
\def\ra{\rangle}
\def\a{\alpha}
\def\b{\beta}

\def\e{\epsilon}

\def\m{\mu}

\begin{document}
\begin{center}
	{\Large \bf Universal superposition of orthogonal states}\\
	\vspace{1cm} 
	Mina Doosti\footnote{doosti\_mina@physics.sharif.edu}, Farzad  Kianvash\footnote{farzadkianvash@gmail.com}, and\\  Vahid  Karimipour\footnote{vahid@sharif.edu}

	\vspace{5mm}
	
	\vspace{1cm} Department of Physics, Sharif University of Technology,\\
	P.O. Box 11155-9161,\\ Tehran, Iran
\end{center}
\vskip 2.5cm


\begin{abstract}                
It is known that no quantum process can produce a predetermined superposition of unknown arbitrary states \cite{Oszmaniec}\cite{Adder}. It has already been shown that with some partial information about the states, one can produce with some probability such superpositions \cite{Oszmaniec}. In general the success probability of these machines, even for orthogonal states, are less than unity. Here we show that there are specific  machines which can produce superpositions of orthogonal qubit states with unit probability. \end{abstract}


\section{Introduction}
Quantum mechanics has continued to bewilder us both with its power for surpassing our classical limitations in doing tasks which are impossible in the classical world, and in showing us limitations in the quantum domain itself for doing tasks which we had taken for granted from the very beginning of its inception. Various tasks in quantum computation and quantum communication, which are now too many and too well-known to list here, \cite{Shor}\cite{Grover}\cite{Gisin}\cite{Giovannetti} are examples of the first aspect, and many no-go theorems like the no-cloning \cite{Wootters}, no-deleting \cite{Pati1}, and recently no-superposition \cite{Oszmaniec} are examples of the second aspect. These are but a sample of many no-go theorems discovered in recent years  \cite{Dieks}\cite{Barnum}\cite{Pati2}\cite{Piani} with many interesting consequences and interrelationships  with for example no-signaling theorem \cite{Ekert}, and uncertainty relations. \\

 The most recent of these theorems which seems to be different from the rest is the no-superposition theorem \cite{Oszmaniec}. It states that there is no universal machine like $\Lambda_{\a,\b}$ which can take two arbitrary states $|\psi\ra$ and $|\phi\ra$ and outputs a fixed superposition of them in the form $|\Psi_{\a,\b}\ra=\a|\psi\ra+\b|\phi\ra.$ This is a striking result in view of the fact that superposition is at the heart of quantum mechanics and many of its intriguing properties, including entanglement and the extraordinary power of quantum computation, stem from this single property. \\
 
 In \cite{Oszmaniec} after proving this no-go theorem for arbitrary input states, the authors proceed to construct a restricted form of superposition machine which can superpose a subclass of states for which we have some prior information. More precisely let $|\chi\ra$ be a state and let $C_1$ and $C_2$ be the following subset of the full Hilbert space $H$ in which $|\psi\ra$ and $|\phi\ra$ are living:
 
 \ba\label{rest}
 C_1&:=&\{|\psi\ra\in H\mid |\la \chi|\psi\ra|^2=c_1 \}\cr
 C_2&:=&\{|\phi\ra\in H\mid |\la \chi|\phi\ra|^2=c_2 \}.
 \ea
Then it is shown in \cite{Oszmaniec} that there is  a universal superposition machine $\Lambda_{\chi,\a,\b}$ abbreviated as $\Lambda$ hereafter, which takes any $|\psi\ra\in C_1$ and any $|\phi\ra\in C_2$ to an un-normalized superposed state 
\be\label{sup1}
|\Psi\ra=\a\frac{\la \chi |\phi\ra}{|\la \chi |\phi\ra|}|\psi\ra+\b\frac{\la \chi |\psi\ra}{|\la \chi |\psi\ra|}|\phi\ra, \ee
where $\a$ and $\b$ are two complex numbers such that $|\a|^2+|\b|^2=1.$ This superposition succeeds with probability 
 
 \be\label{formOzm}
 P_{succ}=\frac{c_1c_2}{c_1+c_2}{{\cal N}_{\Psi}}^2,
 \ee
 where ${\cal N}_\Psi$ is the norm of the state \eqref{sup1}.
  
  Obviously, in addition to the parameters $c_1$ and $c_2$ which characterize the two classes $C_1$ and $C_2$, this success probability depends on which particular states in the two classes are being superposed. Using Cauchy-Schwarz inequality and the condition $|\a|^2+|\b|^2=1$, it is seen that ${{\cal N}_\Psi}^2\leq |\a|^2+|\b|^2+2|\a||\b||\la \psi|\phi\ra|\leq 2$. However the maximum value of $2$ is obtained for the  case  where the two states are parallel and hence their superposition is 
 trivial. For example for two orthogonal states, ${{\cal N}_\Psi}^2=1$ and  hence  
       $P_{succ}=\frac{c_1c_2}{c_1+c_2}$.  Since $c_1$ and $c_2$ are positive numbers  with $0\leq c_1, c_2\leq 1$, the success probability for superposing orthogonal states can never exceed $\frac{1}{2}$ (when $c_1=c_2=1$).\\

 It should be noted that what is really achieved by this superposition machine is not what has initially been sought for, that is a superposition of unknown states with fixed amplitudes $\a$ and $\beta$  and an output state $\a|\psi\ra+\beta|\phi\ra$. In fact, as discussed in detail in \cite{Oszmaniec}, such a machine cannot exist due to the fact that an unphysical phase change like 
 \be
 |\psi\ra\lo e^{i\xi_1} |\psi\ra, \h  |\phi\ra\lo e^{i\xi_2} |\phi\ra, \h
 \ee
 results in a physical change in the output state $\a|\psi\ra+\beta|\phi\ra$ and inevitably the most one can expect of a superposition machine is to produce a state of the form \eqref{sup1}. This form on the one hand cures this unphysical dependence on the phases, and on the other hand, creates superpositions for which only the square of the amplitudes are predetermined to be $|\a|^2$ and $|\b|^2$ and leaves their phases to depend on the initial states.   In passing we note that in \cite{Oszmaniec}, an optical realization of such a machine is also proposed and in \cite{suter},  an explicit NMR realization of such a superposition machine has been reported for qubits. \\ 
 
\begin{figure}[h]\label{fig1}
\centering
\includegraphics[width=13cm,height=6.5cm,angle=0]{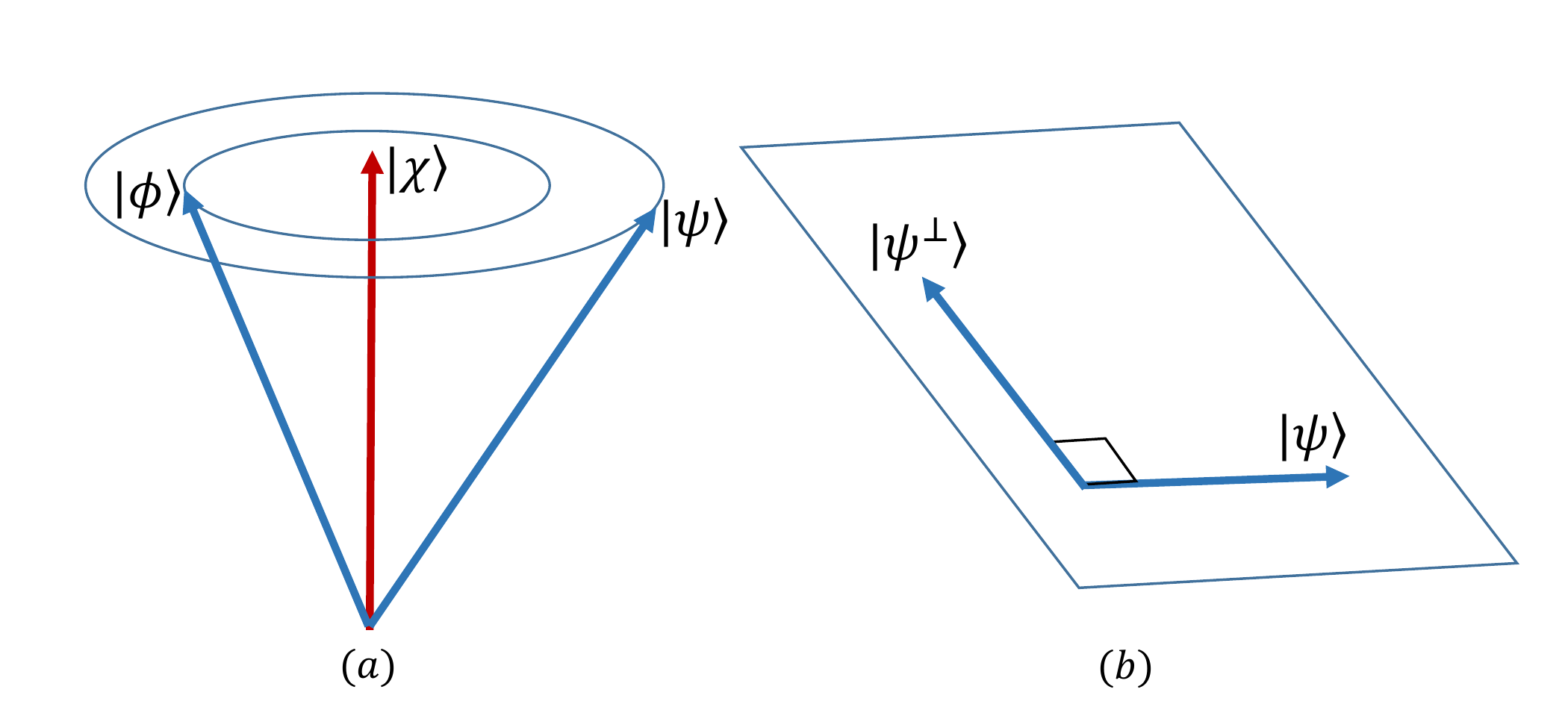}
\caption{a) The class of states which are superposed by the universal machine of \cite{Oszmaniec}. The working of this machine requires that we know the parameters $c_1=|\la \psi|\chi\ra|^2$ and $c_2=|\la\phi|\chi\ra|^2$. Hence, for a fixed set of parameters, the machine superposes only the states on the two cones. For other states, the settings of the machine should change accordingly (See equations 11-14 of \cite{Oszmaniec} and the expression for $P_\m$ therein.) b) The class of states which are superposed in our machine. All the orthogonal states lying on a fixed hyperplane can be superposed with unit probability. For qubits, the hyperplane is the whole Hilbert space, hence no restriction.  }
\end{figure}

In view of these circumstances, i.e. i) the necessity of having partial information about the input states, ii) the form of the output superposed state in which only the moduli of the coefficients are fixed,  and iii) a success probability of less than one for the process, we ask if it is possible from the very beginning to start from a different restricted class of states than \eqref{rest} and construct a superposition machine, which can superpose this new restricted class with a much higher or even unit probability of success. In this paper we show that indeed it is possible to construct such a machine. The restricted class that we start with is the class of arbitrary orthogonal input states, lying on a particular hyperplane, figure (\ref{fig1}). For qubit states, these orthogonal states will be arbitrary, since the hyperplane is the whole Hilbert space.  Therefore we construct a machine $\Lambda_p$ which takes any two orthogonal states $|\psi\ra$ and $|\psi^\perp\ra$  in this hyperplane and outputs a state of the form 
 \be \label{desiredsuperposition}
 |\Psi\ra=\text{{$e^{i\epsilon}$}}(\a|\psi\ra+\beta e^{i\eta}|\psi^\perp\ra),
 \ee
 where now $\epsilon$ and $\eta$ depends on the initial states. The important point is that the success probability of this machine is unity, that is, we always find a superposed state of the above form in the output port of the machine. The difference between our machine and that of \cite{Oszmaniec} is best illustrated in figure \ref{fig1}.  \\
 
 {\bf Remark:} The appearance of a phase factor depending on the initial states in the output state is inevitable. In fact this is the case also in the superposition machine of \cite{Oszmaniec}, \ref{sup1}, where the phase appears through the overlap of the specific states $\phi$ and $\psi$ with $|\chi\ra$, of which only the absolute values are known in the restricted set \ref{rest}. In any superposition machine, only the absolute values of the coefficients $\a$ and $\b$ are fixed in the output. \\

 The structure of this paper is as follows: In section \ref{sec2}, we first consider a machine which outputs a single pure superposition state and calculate the probability of success of this machine. We compare its average probability of success and show that it is higher than  the average probability of success of the machine of \cite{Oszmaniec} when it is restricted to orthogonal states. In section \ref{sec3}, we then discuss a machine which produces a mixture of four output states, all of which are in the form of \eqref{desiredsuperposition} albeit with different phases, where each of them can be selected by measurement of two qubit ancillas. The total probability of the success is now equal to one. We end the paper with a discussion.  \\
  
\section{Orthogonal states superposition with pure output states}\label{sec2}
In this section we introduce a superposition machine for orthogonal input states whose output state is a single pure state.

Let $\ket{\psi}$ and $\ket{\psi^\perp}$ be two arbitrary orthogonal pure states in a two dimensional Hilbert subspace {$H_0$} of the full Hilbert space {$\mathcal{H}$}. The superposing machine is a completely positive map  $\Lambda \in \mathcal{CP}(\mathcal{H}^{\otimes 2},\mathcal{H})$ which acts as follows: 
\begin{equation} \label{eq:first}
	\Lambda_p (\rho \otimes \rho^\perp) \propto \ket{\Psi}\bra{\Psi},
\end{equation}
where
\begin{eqnarray}
	&& \rho = \ket{\psi}\bra{\psi} \cr
	&& \rho^\perp= \ket{\psi^\perp}\bra{\psi^\perp} \crcr
	&&  \ket{\Psi}=\alpha \ket{\psi} +\beta e^{i\eta}\ket{\psi^\perp}
\end{eqnarray}
and $|\alpha|^2 +|\beta ^2|=1$.

Since the output state is pure, the completely positive $\mathcal{CP}$  map is bound to have one Kraus operator, and should be of the form  

\begin{equation} 
\Lambda_p (\rho\otimes \rho^\perp) = K \rho\otimes \rho^\perp K^\dagger\propto |\Psi \ra\la \Psi|,
\end{equation}
 implying that 
\begin{equation}\label{eq:Klinear}
K\ket{\psi} \otimes \ket{\psi^\perp}\propto (\alpha \ket{\psi}+ \beta e^{i\eta}  \ket{\psi^\perp}),
\end{equation}
where $\eta$ is a  function of $\ket{\psi}$ and $\ket{\psi^\perp}$.
Let $\{|0\ra, |1\ra\}$ be an arbitrary basis of $H_0$.  The two states are then parameterized as (with $|x|^2+|y|^2=1$)
\begin{eqnarray} \label{eq:form}
&& |\psi\rangle = \left(\begin{array}{c} x \\ y\end{array}\right) \cr
&& |\psi^\perp\rangle = \left(\begin{array}{c} \overline{y} \\ -\overline{x}\end{array}\right).
\end{eqnarray}
 We will show in the appendix that the desired Kraus operators can have one of the following two forms 

\begin{equation}\label{K1}
K_1 = C
\begin{pmatrix}
  \alpha & 0 & \beta & 0 \\
  0 & 0 & \alpha & \beta
\end{pmatrix},
\end{equation}
or 
\begin{equation}\label{K2}
K_2 = C
\begin{pmatrix}
   -\beta & \alpha & 0 & 0\\
  0 & -\beta & 0 & \alpha
\end{pmatrix},
\end{equation}
where $C$ can be made as large as possible as far as the condition of trace-decreasing $CP$ map i.e. $K_i^\dagger K_i\leq I$ is not violated. Calculation of the greatest eigenvalue of these later matrices shows that we can take $$C=\frac{1}{\sqrt{1+|\a\b|}}$$. \\

 Either of these two solutions will lead to a superposition of two arbitrary orthogonal states. 
To see that it actually does the right task, we check the first solution in detail:

\ba
{K_1}(|\psi\ra\otimes |\psi^\perp\ra)&=&C\begin{pmatrix}
  \alpha & 0 & \beta & 0 \\
  0 & 0 & \alpha & \beta
\end{pmatrix} \left(\begin{array}{c} x\overline{y}\\ -x\overline{x}\\ y\overline{y}\\ -y\overline{x}\end{array}\right)\cr
&=& C \left(\begin{array}{c} \alpha x\overline{y}+\beta y \overline{y}\\ \a y\overline{y} -\beta y \overline{x}\end{array}\right)= C(\a \overline{y} |\psi\ra + \beta y |\psi^\perp\ra)\cr 
&=&C\overline{y}\left[\a  |\psi\ra + \beta e^{i\eta}|\psi^\perp\ra\right],
\ea
where we have used the crucial fact that $\frac{y}{\overline{y}}$ is a phase denoted by $e^{i\eta}$. It is important  to note that this happens only for orthogonal states. In fact it is readily seen that feeding two arbitrary states into this machine leads to a superposition where the coefficients of the two states do not square anymore to $|\a|^2$ and $|\beta|^2$. Note also that the action of this machine can be written in the following compact form

\be \label{superposition}
K_1|\psi\ra\otimes|\psi^\perp\ra=C(\la 0|\psi^\perp\ra) \left(\alpha \ket{\psi}+ \beta \frac{\bra{1}\ket{\psi}}{\bra{0}\ket{\psi^\perp}}  \ket{\psi^\perp}\right).
\ee

Again we note that in any basis, only for orthogonal states the factor $\frac{\la 1|\psi\ra}{\la 0|\psi^\perp\ra}$ is a pure phase. 
Similar calculation shows that

\be \label{superposition2}
K_2|\psi\ra\otimes|\psi^\perp\ra=C(\la 1|\psi^\perp\ra) \left(\alpha \ket{\psi}- \beta \frac{\bra{0}\ket{\psi}}{\bra{1}\ket{\psi^\perp}}  \ket{\psi^\perp}\right).
\ee

The above two machines are related in the following way where we have used the temporary superscripts on $K_1$ and $K_2$ to denote the coefficients of the superposition machines which they represent:
\be
{K_1}^{\a,\b}|\psi\ra\otimes |\psi^\perp\ra={K_2}^{\a\lo \b,\b\lo -\a}|\psi^\perp\ra\otimes |\psi\ra.
\ee
Let us now calculate the probability of the above two machines, where the superscript 1 or 2, denotes the type of machine. Using the parameterization $x=\cos\frac{\theta}{2}, y=\sin\frac{\theta}{2} e^{i\phi}$, we find

\be
{P^1}_{success}=|K_1|\psi\ra\otimes|\psi^\perp\ra|^2=C^2|\la 0|\psi^\perp\ra|^2=C^2 \cos^2\frac{\theta}{2}.
\ee

and

\be
{P^2}_{success}=|K_2|\psi\ra\otimes|\psi^\perp\ra|^2=C^2|\la 1|\psi^\perp\ra|^2=C^2 \sin^2\frac{\theta}{2}.
\ee

which shows the complementary roles of the machines in superposing different states. As an example note that the machine 1, superposes the states $|\psi\ra=|1\ra$  and $|\psi^\perp\ra=|0\ra$  with probability equal to $C^2$ but does not superpose at all the states $|\psi\ra=|0\ra$  and $|\psi^\perp\ra=|1\ra$. The machine 2 does the converse job. Both machines superpose the two states $|\psi\ra=|\pm\ra$ and $|\psi^\perp\ra=|\mp\ra$ with $P_{succ}= \frac{C^2}{2}$, i.e. more explicitly from \eqref{superposition} and \eqref{superposition2} we find:
\be
K_1|+\ra\otimes |-\ra=K_2|-\ra\otimes|+\ra=\frac{C}{\sqrt{2}}(\a|+\ra+\b|-\ra).
\ee
The average probability for both machines, averaged over the Bloch sphere  is given by 
\be
	\la P^{1,2}_{succ}\ra=\int P^{1,2}_{succ} d\Omega
	=\frac{C^2}{2} = \frac{1}{2(1+|\a||\b|)}.
\ee
This probability depends on $|\a|$ and $|\b|$, but is always greater than $\frac{1}{3}.$
A comparison with the average probability of the general machine of \cite{Oszmaniec} is in order now. 
To this end we quote the basic relations of \cite{Oszmaniec} for ease of reference. The success probability for any two states is given by 
\be
P_{succ}=\frac{c_1c_2}{c_1+c_2}{{\cal N}_{\Psi}}^2,
 \ee
where
\be
c_1=tr(P_{\chi}P_{\psi}),\h c_2=tr(P_{\chi}P_{\phi}),
\ee 
   and ${\cal N}_{\Psi}$ is found from \cite{Oszmaniec} to be 
     \be
    {{\cal N}_{\Psi}}^2=1+2\cdot  Re\left(\overline{\a}\b \frac{Tr(P_\chi P_\psi P_\phi)}{\sqrt{Tr(P_\chi P_\psi)Tr(P_\chi P_\phi)}}\right).
   \ee 
  Consider now two orthogonal states 
  \be
  P_\psi=\frac{1}{2}(I+{\bf n}\cdot \sigma),\h  P_\phi=\frac{1}{2}(I-{\bf n}\cdot \sigma),
  \ee 
   and an arbitrary state 
   \be
   P_\chi =\frac{1}{2}(I+{\bf s}\cdot \sigma),
   \ee
   where ${\bf n}$ and ${\bf s}$ are unit vectors. Then we find that $\mathcal{N}_\Psi=1$ (since $P_\psi P_{\psi^\perp}=0$), and 
   \be
   c_1=\frac{1}{2}(1+{\bf n}\cdot {\bf s}), \h   c_2=\frac{1}{2}(1-{\bf n}\cdot {\bf s}).
   \ee
  This leads to the success probability 
  \be
  P_{succ}=\frac{1}{4}(1-({\bf n}\cdot {\bf s})^2)
  \ee
   which of course depends on the initial state and the overlap state $|\chi\ra$, but is indepdendent of the values of $\a$ and $\beta$. Averaged over the Bloch sphere, this gives 
   \be
    \la P_{succ}\ra=\int \frac{1}{4}(1-({\bf n}\cdot {\bf s})^2)d\Omega=\frac{1}{4}(1-\frac{1}{3})=\frac{1}{6}.
   \ee
   It is clearly seen that the performance of the general purpose superposition machine is much lower than the special purpose superposition machine. In the next section we improve our machine so that it will produce such a superposition with unit probability.

       \section{Orthogonal states superposition with mixed  output states} \label{sec3}
      Our freedom in choosing $\eta$ in Eq. (\ref{eq:Klinear}) helps us to design even better superposition machines, machines with $P_{succ}=1$. Consider a $\mathcal{CPT}$ map $\Gamma$ which acts as follows
\begin{equation}
	\Gamma(\rho\otimes\rho^\perp)=\sum_{j}^dA_j\rho\otimes\rho^\perp A_j^\dagger
\end{equation}
where $ \sum\limits_{j}^{d}A_j^\dagger A_j=I,
$
and $d$ is the number of Kraus operators.\\
If the Kraus operators satisfy
\begin{equation}\label{probability=1}
	A_j\ket{\psi} \ket{\psi^\perp}= C_j (\alpha \ket{\psi}+ \beta \e^{i\eta_j}  \ket{\psi^\perp})=C_j\ket{\Psi^j},
\end{equation}
then the output of the quantum channel is $\ket{\Psi^j} \quad j\in\{1,2,..,d\}$ with probability $\abs{C_j}^2$. In all of these states, the absolute values of probability amplitudes are $\abs{\alpha}$ and $\abs{\beta}$. 
Therefore, Eq. (\ref{probability=1})  guarantees that the superposition machine's output has the proper form.\\

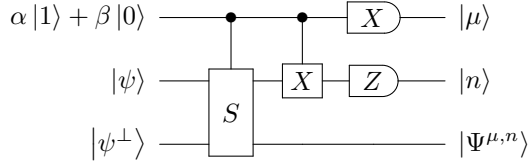
\begin{figure}[t]
	\Qcircuit @C=.7em @R=.4em @! {
		&&&&&\lstick{\alpha\ket{1}+\beta\ket{0}} & \ctrl{1} & \ctrl{1} & \measureD{X} & \rstick{\ket{\mu}} \qw \\
		&&&&&\lstick{\ket{\psi}} & \multigate{1}{S} & \gate{X} &\measureD{Z} &  \rstick{\ket{n}} \qw\\
		&&&&&\lstick{\ket{\psi^\perp}} & \ghost{S} & \qw & \qw & \rstick{\ket{\Psi^{\mu,n}}} \qw
	}
	\caption{The superposing machine for orthogonal states. The inputs are an ancillary qubit and any two orthogonal states.}
	\label{fig22}
\end{figure}
 Consider the quantum circuit depicted in figure \ref{fig22} , where $X$ is the Pauli gate $X=\left( \begin{array}{cc} 0 & 1 \\
1 & 0 \end{array} \right)$ and $S$ is the swap operator which interchanges its two inputs. At  the end of the circuit, we perform two measurements on the first and second qubits, in the $X$ and $Z$ bases  i.e. in the bases  $\ket{\pm}=\frac{\ket{0}\pm\ket{1}}{\sqrt{2}}$ and $\ket{0},\ket{1}$  respectively. There are four possible outcomes for the measurements, and there are four corresponding Kraus operators as following
\begin{align}
	A_{\m ,n}=(\ket{\m}\bra{\m}\otimes \ket{n}\bra{n} \otimes \mathbb{I})\circ U
\end{align}
where $\ket{\m}=\ket{\pm}$ and $\ket{n}=\ket{0},\ket{1}$, and $U$ is the unitary operations performed in the circuit
\begin{align}
	U=&(\ket{0}\bra{0}\otimes \mathbb{I}\otimes \mathbb{I} + \ket{1}\bra{1}\otimes X \otimes \mathbb{I} )\circ\cr
	&(\ket{0}\bra{0}\otimes \mathbb{I}\otimes \mathbb{I} + \ket{1}\bra{1} \otimes S).
\end{align}
Defining $|anc\ra:=\a|0\ra+\b|1\ra$, simple calculation shows that 
\be
U|anc\ra\otimes|\psi\ra\otimes|\psi^\perp\ra=\a|1\ra \otimes X|\psi^\perp\ra \otimes|\psi\ra+\beta |0\ra\otimes|\psi\ra \otimes|\psi^\perp\ra
\ee
 from which we obtain the final state to be 
\begin{equation}
	A_{\m,n}U|anc\ra|\psi\ra|\psi^\perp\ra =\a\la \m|1\ra\la n|X|\psi^\perp\ra|\psi\ra+\beta \la \m|0\ra\la n|\psi\ra|\psi^\perp\ra.
\end{equation}
Since $\la \m|1\ra=\frac{\m}{\sqrt{2}}$ and $\la \m|0\ra=\frac{1}{\sqrt{2}}$, this leads to 
\be
A_{\m,n}U|anc\ra|\psi\ra|\psi^\perp\ra =\frac{\m}{\sqrt{2}}\la n|X|\psi^\perp\ra \left(\a|\psi\ra + \beta\mu \frac{\la n|\psi\ra}{\la n|X|\psi^\perp\ra}|\psi^\perp\ra\right),
\ee
where again we note that in any basis the factor $\frac{\la n|\psi\ra}{\la n|X|\psi^\perp\ra}$ is a pure phase and hence the superposition machine outputs the following states  

\begin{align}\label{probability=1ex}
	\ket{\Psi^{\m,n}}=\alpha \ket{\psi} + \beta e^{i\eta_{\m,n}} \ket{\psi^\perp}
\end{align}
where $e^{i\eta_{\m,n}}=\m\frac{\bra{n}\ket{\psi}}{\bra{n}X\ket{\psi^\perp}}$ and the probabilities given by  
\begin{equation}
	P_{\m,n}=\frac{1}{2}\abs{\bra{n}X\ket{\psi^\perp}}^2
\end{equation}
Note that Eq. (\ref{probability=1ex}) is in the form of (\ref{probability=1}). Although the amplitudes of the superposition in $\ket{\Psi^{\m,n}}$ are unknown and dependent on $\m,n$, but their absolute values are $\abs{\alpha}$ and $\abs{\beta}$. As can be seen in table \ref{table:questions}, all of the outcomes of the superposition machine are in the desired form. Consequently, the total probability of success for this superposition machine is equal to one.
	\begin{table}
	
		\centering
	\begin{tabular}{||c c c c||} 
		\hline
		X & Z & $\ket{\Psi^{\mu,n}}$ & $P_{\mu,n}$  \\ 
		\hline\hline
		$\mu=+$ & $n=0$ & $ \alpha\ket{\psi}-\beta e^{-i\phi}\ket{\psi^\perp}$& $\frac{1}{2}\sin[2](\theta)$ \\ 
		\hline
		$\mu=+$ & $n=1$ & $ \alpha\ket{\psi}+\beta e^{i\phi}\ket{\psi^\perp}$& $\frac{1}{2}\cos[2](\theta)$ \\
		\hline
		$\mu=-$ & $n=0$ & $ \alpha\ket{\psi}+\beta e^{-i\phi}\ket{\psi^\perp}$ & $\frac{1}{2}\sin[2](\theta)$ \\ 
		\hline
		$\mu=-$ & $n=1$ & $ \alpha\ket{\psi}-\beta e^{i\phi}\ket{\psi^\perp}$ &$\frac{1}{2}\cos[2](\theta)$ \\
		\hline
	\end{tabular}
	\caption{Different output states of the machine (\ref{fig22}) and their corresponding probabilities for obtaining them. }
			\label{table:questions}
\end{table}

\section{Relation with cloning}
It is an intriguing question whether or not there is any relation between the no-cloning theorem and the no-superposition theorem. As stated in \cite{} it appears that these two no-go theorems have quite different characters and one cannot conclude one from the other. For example one can imagine senarios where perfect cloning a perfect deleting machines exist and then use them in conjunction with the orthogonal superposition machine of the present paper to prepare superposition of arbitrary states. More concretely given two states $|\phi\ra$ and $|\psi\ra$, one prepares perfect clones $|\phi\ra^{\otimes N}$ and $|\psi\ra^{\otimes N}$, which for large $N$ are orthogonal and then feeds them to the superposition machine proposed in this paper which outputs a state of the form $|\Psi\ra=\a |\phi\ra^{\otimes N}+\b|\psi^{\otimes N}\ra$. Deleting the first $N-1$ copies now produces the desired superposed state $\a|\phi\ra+\b|\psi\ra$ on the last qubit. However a closer look at the deleting machine shows that this is impossible. In fact a deleting machine which deletes the first $N-1$ qubits works as follows:

\ba
|\phi^{\otimes N}\ra&\lo& |0^{\otimes N}\ra\otimes |\phi\ra\otimes |A_\phi\ra\cr
|\psi^{\otimes N}\ra&\lo& |0^{\otimes N}\ra\otimes |\psi\ra \otimes |A_\psi\ra,
\ea
where $A_\phi$ and $A_\psi$ are the states of the deleting machine which are nearly orthogonal due to the unitarity of the operation. This then means that the output state of the $N-$th qubit is a mixed state $|\a|^2 |\phi\ra\la \phi|+|\b|^2 |\psi\ra\la \psi|$ and not a superposition. In this regard we can also mention a recent paper \cite{unif} which tries to put different no-go theorems like the no-cloning, no-deleting and no-superposition theorems in a unified framework. However this doesn't mean that a profound basis is found which connects these no-go theorems, rather  different no-go theorems are combined to  form new bigger no-go theorems. A sample of the theorems proved in \cite{unif} is the following: If $|\phi\ra$
 is a fixed state and $|\psi\ra$ is an arbitrary state both belonging to a Hilbert space $H$, then there is no map ${\cal F}: H^{\otimes k} \lo H^{\otimes n}$ such that ${\cal F}({\rho_{\psi}}^{\otimes k})={\rho_\chi}^{\otimes n}$, where $|\chi\ra\propto \a |\psi\ra+\b |\phi\ra$. As it is seen, this is a combination of the no-cloning and the no-superposition theorems which while valuable in itself, does not solve the open question of the relation between the two theorems. 

\section{Summary}
The no-superposition theorem \cite{Oszmaniec} forbids superposition of any two arbitrary states and only allows superposition of states for which partial information, i.e. the overlaps of the two states with a fixed state is known \ref{rest} is available. We have shown that for orthogonal states lying on a hyperplane, such superposition is possible with unit probability. In all these machines, including the one proposed in \cite{Oszmaniec}, the appearance of a phase which depends on the input states is inevitable. In fact given two states $|\phi\ra$ and $|\psi\ra$, a superposition machine outputs a state of the form $\a|\psi\ra+e^{i\xi}\b |\psi\ra$ where $\a$ and $\b$ are fixed by the machine and $\xi$ depends on the input states. It is the appearance of this phase, which has enabled us to propose a process by which orthogonal states on a hyperplane can be superposed with unit probability, figure \ref{fig22} and  table \ref{table:questions}. As we have discussed the relation between the no-cloning theorem and the no-superposition theorem remains open.

\appendix
\section{Finding the Kraus Operators } \label{Appendix A}
In this appendix we sketch the steps for finding the form of Kraus operators $\ref{K1}$ and $\ref{K2}$.
 We require a single Kraus operator K satisfying following relation
\begin{equation}\label{eq:AppendixK}
K\ket{\psi} \ket{\psi^\perp}= \delta(\theta, \phi)(\alpha \ket{\psi}+ \beta \e^{i\eta}  \ket{\psi^\perp}),
\end{equation}
in which $\delta(\theta, \phi)$ and $\eta(\theta,\phi)$ are functions of parameters of input states. Since $K$ is an operator which is non trivial only on a hyperplane of the total Hilbert space , it can be written as a 2x4 matrix. Let us parameterize the states as  
$|\psi\ra=\left(\begin{array}{c} \sin{\frac{\theta}{2}} \\ \cos{\frac{\theta}{2}} e^{i\phi}\end{array}\right)$ and $   |\psi^\perp\ra=\left(\begin{array}{c} \cos{\frac{\theta}{2}}\\ -\sin{\frac{\theta}{2}} e^{i\phi}\end{array}\right)$, and the K operator as 
\begin{eqnarray} \label{appendix2}
&&
\begin{pmatrix}
  x & y & z & w \\
  x' & y' & z' & w'
\end{pmatrix}
\begin{pmatrix}
\sin{\frac{\theta}{2}} \cos{\frac{\theta}{2}} \\
-\sin^2{\frac{\theta}{2}} e^{i\phi} \\
\cos^2{\frac{\theta}{2}} e^{i\phi} \\
-\sin{\frac{\theta}{2}} \cos{\frac{\theta}{2}} e^{2i\phi} 
\end{pmatrix} \cr
&&= \delta(\theta, \phi)
\begin{pmatrix}
\alpha \sin{\frac{\theta}{2}} + \beta e^{i\eta} \cos{\frac{\theta}{2}} \\
\alpha e^{i\phi}\cos{\frac{\theta}{2}} - \beta e^{i\eta} e^{i\phi} \sin{\frac{\theta}{2}}
\end{pmatrix}
\end{eqnarray}

Operating the K matrix leads to the following two equations:
\ba\nonumber
&(x-w\ e^{2e^{i\phi}})\sin\frac{\theta}{2} \cos\frac{\theta}{2} + e^{i\phi}(z\cos^2\frac{\theta}{2}-y\sin^2\frac{\theta}{2})=\delta[\a \sin\frac{\theta}{2} + \beta e^{i\eta}\cos\frac{\theta}{2}]\cr
&(x'-w'\ e^{2e^{i\phi}})\sin\frac{\theta}{2} \cos\frac{\theta}{2} + e^{i\phi}(z'\cos^2\frac{\theta}{2}-y'\sin^2\frac{\theta}{2})=\delta[\a \cos\frac{\theta}{2} - \beta e^{i\eta}\sin\frac{\theta}{2}]\ e^{i\phi}.
\ea
 
To find a simple solution for the K operator, we use the fact that it should be independent of the parameters of the input states, and take the function $\delta(\theta, \phi)$ to be a first degree function of  $\sin{\frac{\theta}{2}}$ and $\cos{\frac{\theta}{2}}$, while assuming for simplicity that $\eta$ is  independent of $\theta$. 
Therefore we find
\begin{equation}\label{eq:AppendixCform}
\delta(\theta, \phi) = a(\phi) \sin{\frac{\theta}{2}} + b(\phi) \cos{\frac{\theta}{2}}
\end{equation}
Equating the coefficients of $\sin^2\frac{\theta}{2}$,  $\cos^2\frac{\theta}{2}$ and $\sin\frac{\theta}{2}\cos\frac{\theta}{2}$ on both sides leads to 
\be\label{AA1}
e^{i\phi}z=b(\phi)\beta e^{i\eta} \  \ , \ \  z'=b(\phi)\a,
\ee
\be\label{AA2}
-e^{i\phi}y=a(\phi)\a, \  \ , \ \   y'=a(\phi)\beta e^{i\eta},
\ee
and
\be\label{AA3}
(x-w e^{2i\phi})=a(\phi)\beta e^{i\eta}+b(\phi) \a\  \ , \ \ (x'-w' e^{2i\phi})=a(\phi)\a e^{i\phi}- b(\phi)\beta
e^{i(\eta+\phi)}.
\ee

We now use the fact that the parameters $x,y,z,w$ and $x',y',z',w'$ are independent of $\phi$, but the parameters $a,b$ and $\eta$ can depend on $\phi$. Therefore we find from the left hand side 
equation of (\ref{AA2}) and the right hand side of equation (\ref{AA1}) the following two relations respectively:

\be
a(\phi)=-\frac{y}{\a} e^{i\phi}\ \ \ \  b(\phi)=\frac{z'}{\a}.
\ee

Inserting these in the remaining two equations of (\ref{AA1}) and (\ref{AA2}), we find
\be
y'=-\frac{\beta}{\a}\ y\ e^{i(\phi+\eta)},
\ee
and
\be
z'=\frac{\a}{\b}\ z\ e^{i(\phi-\eta)}.
\ee

Demanding that $y,y'$ and $z,z'$ be indepdendent of $\phi$, now leads to one of the following two solutions:\\

{\bf Solution 1:} $\eta=\phi$, which leads to 
\be
z'=\frac{\a}{\b}z,\ \ \ y=y'=a(\phi)=0.
\ee
Inserting the values $a(\phi)=0$ and $b(\phi)=\frac{z'}{\a}$ from (\ref{AA1}) into (\ref{AA3}) leads to 
\be
x-we^{2i\phi}=z',\ \ \ \  x'-w'e^{2i\phi}=-\frac{\beta}{\a}z' e^{2i\phi},
\ee
the solution of which is  $\omega = 0\ \ \ , x=z',\ \ \ x'=0, $ and $ w'=\frac{\beta}{\a}z' $. This leads to the form $K_1$.\\

{\bf Solution 2:} $\eta=-\phi$, which leads to 
\be
y'=-\frac{\beta}{\a}y,\ \ \ z=z'=b(\phi)=0.
\ee
Inserting the values $b(\phi)=0$ and $a(\phi)=-\frac{y}{\a}e^{i\phi}$ in (\ref{AA3}), leads to 
\be
x-we^{2i\phi}=-\frac{\beta}{\a}y,\ \ \ \  x'-w'e^{2i\phi}=-y e^{2i\phi},
\ee
the solution of which is  $\omega = 0\ \ \ , x=-\frac{\beta}{\a}y,\ \ \ x'=0, $ and $ w'=y. $ This leads to the form $K_2$.\\

\end{document}